 \documentclass[final,5p,times,twocolumn]{elsarticle}

\usepackage{amssymb}

 \usepackage{amsmath}
 \usepackage{color}
 
\journal{Particle Astrophysics and Cosmology B}

\begin{document}

\begin{frontmatter}



\author[UPV]{S.~Adri\'an-Mart\'inez}
\author[Colmar]{A.~Albert}
\author[UPC]{M.~Andr\'e}
\author[Erlangen]{G.~Anton}
\author[UPV]{M.~Ardid}
\author[CPPM]{J.-J.~Aubert}
\author[APC]{T.~Avgitas}
\author[APC]{B.~Baret}
\author[IFIC]{J.~Barrios-Mart\'{\i}}
\author[LAM]{S.~Basa}
% ...

\fntext[tagiuf]{\scriptsize{Institut Universitaire de France, 75005 Paris, France}}
\fntext[tag:2]{\scriptsize{Also at INFN-Bari}}
\fntext[tag:3]{\scriptsize{Also at APC}}

\address[UPV]{\scriptsize{Institut d'Investigaci\'o per a la Gesti\'o Integrada de les Zones Costaneres (IGIC) - Universitat Polit\`ecnica de Val\`encia. C/  Paranimf 1 , 46730 Gandia, Spain.}}
\address[Colmar]{\scriptsize{GRPHE - Universit\'e de Haute Alsace - Institut universitaire de technologie de Colmar, 34 rue du Grillenbreit BP 50568 - 68008 Colmar, France}}
\address[UPC]{\scriptsize{Technical University of Catalonia, Laboratory of Applied Bioacoustics, Rambla Exposici\'o,08800 Vilanova i la Geltr\'u,Barcelona, Spain}}
\address[Erlangen]{\scriptsize{Friedrich-Alexander-Universit\"at Erlangen-N\"urnberg, Erlangen Centre for Astroparticle Physics, Erwin-Rommel-Str. 1, 91058 Erlangen, Germany}}
\address[CPPM]{\scriptsize{Aix-Marseille Universit\'e, CNRS/IN2P3, CPPM UMR 7346, 13288 Marseille, France}}
\address[APC]{\scriptsize{APC, Universit\'e Paris Diderot, CNRS/IN2P3, CEA/IRFU, Observatoire de Paris, Sorbonne Paris Cit\'e, 75205 Paris, France}}
\address[IFIC]{\scriptsize{IFIC - Instituto de F\'isica Corpuscular, Edificios Investigaci\'on de Paterna, CSIC - Universitat de Val\`encia, Apdo. de Correos 22085, 46071 Valencia, Spain}}
\address[LAM]{\scriptsize{LAM - Laboratoire d'Astrophysique de Marseille, P\^ole de l'\'Etoile Site de Ch\^ateau-Gombert, rue Fr\'ed\'eric Joliot-Curie 38,  13388 Marseille Cedex 13, France}}
% ...
\address[Bari]{\scriptsize{INFN - Sezione di Bari, Via E. Orabona 4, 70126 Bari, Italy}}
% ...

\title{Limits on Dark Matter Annihilation in the Sun using the ANTARES Neutrino Telescope}

\begin{abstract}

A search for muon neutrinos originating from dark matter annihilations in the
Sun is performed using the data recorded by the ANTARES neutrino
telescope from 2007 to 2012. In order to obtain the best possible
sensitivities to dark matter signals, an optimisation of the event
selection criteria is performed taking into account the background
of atmospheric muons, atmospheric neutrinos and the energy spectra of the
expected neutrino signals. No significant excess over the background
is observed and $90\%$ C.L. upper limits on the neutrino flux, the
spin--dependent and spin--independent WIMP-nucleon cross--sections are
derived for WIMP masses ranging from $ \rm 50$ GeV to $\rm 5$ TeV
for the annihilation channels $\rm WIMP + WIMP \to b \bar b, W^+ W^-$
and $\rm \tau^+ \tau^-$.

\end{abstract}


\end{frontmatter}


\section{Introduction}
\label{Intro}

A number of independent observations in cosmology and astrophysics point to the existence of large amounts of non--baryonic matter in the Universe ~\cite{DM_rev,darkmatter}. These observations indicate that there is approximately five times more of this dark matter than of ordinary baryonic matter.

A well-motivated hypothesis is that dark matter is composed of weakly interacting massive particles (WIMPs) that form halos in which galaxies are embedded. There are different candidates for these WIMPs, amongst which, those provided by supersymmetric models are currently the focus of the attention of a large variety of searches. In the case of the minimal supersymmetric extension of the Standard Model (MSSM), the lightest new particle is stable due to the conservation of a quantum number, the R-parity, that prevents its decay to ordinary particles \cite{susyprimer}. If this lightest supersymmetric particle is also electromagnetically neutral, it is a natural WIMP candidate for dark matter. This lightest particle can annihilate into pairs of standard model particles. Neutrinos, in particular, are the final product of a large variety of decay processes, being therefore a good candidate for an indirect search for dark matter. WIMPs tend to accumulate in celestial objects due to scattering with ordinary matter and the gravitation pull of these objects. This is why indirect searches for dark matter concentrate on massive astrophysical bodies such as the Earth, the centre of our Galaxy, galaxy clusters or, as in this case, the Sun.

In this letter, an indirect search for neutrinos coming from WIMP annihilations in the Sun is presented, using data recorded by the ANTARES neutrino telescope from 2007 to 2012. Different quality cuts on the data have been used to reduce the atmospheric background and optimise the sensitivity of the analysis. Sensitivities to the signal neutrino flux, $\Phi_{\nu}$, and the spin--dependent and spin--independent WIMP-nucleon cross--sections, $\mathrm \sigma_{\text{SD}}^{\text{p}}$ and $\mathrm \sigma_{\text{SI}}$, are derived using three different annihilation channels. 

\section{The ANTARES neutrino telescope}
\label{antares}

The ANTARES detector \cite{antares,positioning} is an undersea neutrino telescope anchored 2475 m below the surface of the Mediterranean Sea and 40 km offshore from Toulon (France) at $42^\circ 48^\prime$ N and $6^\circ 10^\prime$ E. ANTARES consists of 12 detection lines with 25 storeys per line and 3 optical modules with $\mathrm10^{\prime \prime}$ photomultipliers per storey. The detection lines are 450 m long and 60-75 m apart horizontally. Data taking started in 2007, when the first five lines of ANTARES were installed. The detector installation was completed in May 2008.

The main channel through which neutrinos are detected is via the muons produced from high--energy muon neutrinos interacting inside, or in the vicinity of, the detector. These muons move at relativistic velocities and induce the emission of Cherenkov light that is then detected by the optical modules. In this analysis, only muon neutrinos detected this way will be considered. In the following any mention of `neutrinos` will refer to muon neutrinos and muon antineutrinos.

The flux of atmospheric muons from above the detector comprises the largest part of the background, with fluxes several orders of magnitude larger than any expected signal. In order to reduce the number of atmospheric muons, a cut on the elevation of reconstructed muon tracks is applied, ensuring that only events that have been reconstructed as upgoing are used. Since muons cannot cross the entire Earth, this cut rejects all atmospheric muons except for a small fraction of misreconstructed events. The atmospheric neutrinos represent the irreducible background for this search.

Atmospheric neutrinos from 10~GeV to 20~TeV are generated in the simulation using the standard ANTARES simulation chain \cite{genhen,MUPAGE,km3,transmission,OM,biofouling}. 

The expected neutrino energy spectra from WIMP annihilations in the Sun are calculated with the WIMPSim simulation package \cite{WIMPSim}. The code takes into account the absorption of neutrinos in the solar plasma and the neutrino oscillation inside the Sun and on their way from the Sun to the detector. Neutrino spectra are calculated for 15 WIMP masses ranging from 50~GeV to 5~TeV and three annihilation channels assuming a branching ratio of 100\%:

\begin{equation}
\rm WIMP + WIMP \to b \bar{b}, \tau^+ \tau^-, W^+ W^- . \label{channels}
\end{equation}

As shown in \cite{Ant_dmsun}, the distribution of the number of muon neutrinos arriving at the Earth per pair of WIMPs self-annihilating in the Sun's core provides hard spectra for the $\tau^+ \tau^-$ and $W^+ W^-$ and a soft spectrum for $\rm b \bar{b}$. Limits calculated for dark matter candidate models will lie between these three channels, depending on their branching ratios. The energy spectrum of each channel (see Figure 2 in \cite{Ant_dmsun}) is used to calculate the acceptance for the particular annihilation channel in Equation \ref{channels}. The acceptance is the neutrino effective area convoluted with the energy spectrum corresponding to a given WIMP mass (see Section \ref{method}).

Two reconstruction algorithms are used in this paper. The first one is based on the minimisation of a $\chi^2$-like quality parameter, Q, of the reconstruction which uses the difference between the expected and measured times of the detected photons, taking into account the effect of light absorption in the water\cite{BBFit}. The second algorithm consists of a multistep procedure to fit the direction of the muon track by maximising a likelihood ratio, $\Lambda$, which describes the quality of the reconstruction \cite{AAFit_official}. In addition to the $\Lambda$ parameter, the uncertainty of the muon track angle, $\beta$, is used for the track selection. These two algorithms are respectively called here QFit and $\Lambda \text{Fit}$. In order to reach the best efficiency of reconstruction in the entire neutrino energy range QFit is used for muon events reconstructed in a single detection line (single-line events), and $\mathrm \Lambda \text{Fit}$ for muon events reconstructed on more than one detection line (multi-line events).

Extensive comparisons between data and simulations have been made elsewhere~\cite{Ant_dmgc}. 

\section{Analysis strategy}
\label{method}

The search for WIMP annihilation in the Sun is performed based on a maximum-likelihood analysis method. The maximisation of this likelihood function, which is fed with the known information about the characteristics of the expected background and signal, provides an estimate of the amount of signal in the data. The extended likelihood function used for $\mathrm\Lambda \text{Fit}$ is

\begin{equation}
\begin{split}
 \cal{L} \rm (n_s) = & e^{- (n_s+N_{\text{bg}})} \prod_{i = 1} ^{N_{tot}} \left( n_s S(\psi_i,N_{hit,i},\beta_i)
+N_{\text{bg}}B(\psi_i,N_{hit,i},\beta_i) \right),  \label{lik1}
\end{split}
\end{equation}

\noindent where $ N_{\text{bg}}$ is the expected number of background events, $ N_{\text{tot}}$ is the total number of reconstructed events, $n_s$ (the variable that changes during the maximisation process) is the number of signal events in the likelihood function, $S$ and $B$ are functions that calculate the likelihood of an event to be either signal or background, $\psi_i$ is the angular distance of the $i$-th event to the Sun, $\rm N_{hit,i}$ is the number of hits used in the reconstruction of the $i$-th event, which is used as an energy estimate and $ \beta_i$ is the value of the angular error estimate for the $i$-th event. $S$ is calculated from the simulation and $B$ is calculated from time--scrambled data. 

For the QFit analysis the likelihood function looks different since for that analysis only single--line events have been used. For these events the azimuth angle can not be determined, so that the difference between the zenith angle of the events and the Sun has to be used instead of $\psi$:

\begin{equation}
\begin{split}
 \cal{L} \rm (n_s) = & e^{- (n_s+N_{\text{bg}})} \prod_{i = 1} ^{N_{tot}} \left( n_s \bar S(\theta_i,\bar{N}_{hit,i},Q_i)
+N_{\text{bg}}\bar B(\theta_i,\bar{N}_{hit,i},Q_i) \right) ,   \label{lik2}
\end{split}
\end{equation}

\noindent where $\bar{N}_{hit,i}$ is the number of hits summed up per storey used for the reconstruction and $\theta_i$ is the difference in zenith angle between the $i$-th event and the Sun. $\bar S$ and $\bar B$ are analogous to $S$ and $B$ in the likelihood function used for the $\mathrm \Lambda\text{Fit}$ data.

The angular resolution, which is used in $S$, is limited by the kinematic angle between neutrino and outgoing muon \cite{Ant_dmgc}. 

In this analysis a blinding protocol is applied for optimising the event selection. Blinding is achieved by using simulations to calculate the sensitivities, and time--scrambled data for calculating the background estimate. 

In order to compute sensitivities and limits, $10^4$   {\it pseudo--experiments} are performed for each combination of WIMP mass, annihilation channel and reconstruction strategy and for each considered value of $n_s$ ($n_s \in \{ 0,1,2... 20 \}$). In a pseudo--experiment, a random distribution of background events is simulated according to the features of the recorded data by randomising the right ascension of the events. Simulated signal events are introduced into these pseudo--experiments. These events are generated using the PSF and the signal characteristics for a given reference flux, which are also used in the likelihood function. For each pseudo--experiment, $\rm n_s$ is varied to maximise the likelihood function (when $\rm n_s = n_{max}$). The test statistic (TS) is then calculated as

\begin{equation}
\rm TS =  log_{10}\left(\frac{{\cal L}(n_{max})}{{\cal L}(0)}\right).
\end{equation}

Distributions of TS values are generated for different numbers of injected signal events. The overlap of TS distributions with inserted signal events and the  TS distribution with only background is a measure of the likelihood to mistake pure background for an event distribution with a certain amount of signal in it. From this, the 90\% C.L. sensitivities in terms of detected signal events, $\mu_{90\%}$, are obtained using the Neyman method for generating limits~\cite{Neyman}. The so-defined $\mu_{90\%}$ quantity corresponds to the lowest number of signal events so that 90\% of pseudo-experiments provide  TS values above the median of the TS distribution of the pure background case.

Event selection consists of cuts on the quality parameters $\Lambda$ and Q of the two reconstructions that are used in this analysis. These cuts are optimised with respect to the sensitivities (i.e. the model rejection factor). The optimum cuts for the relevant mass ranges are  $\Lambda > -5.4$ and $\beta < 1^{\circ}$ for $\mathrm \Lambda\text{Fit}$  and $Q < 0.8$ for the QFit analysis. 

The sensitivities in terms of neutrino fluxes are calculated using the acceptance, defined as

\begin{equation}
\begin{split}
 \mathcal{A}^j(M_{\text{WIMP}}) =  & \int_{E_{th}} ^{M_{\text{WIMP}}} 
A^j_{\text{eff}}(E_{\nu_\mu}) \left.
\frac{d\Phi_{\nu_\mu}}{dE_{{\nu}_{\mu}}}\right|_{ch} dE_{\nu_\mu} \cdot T_{\text{eff}}^j\\
& + \int_{E_{th}} ^{M_{\text{WIMP}}} A^j_{\text{eff}}(E_{\bar \nu_{ \mu}}) 
\left.
\frac{d\Phi_{\bar \nu_{ \mu}}}{dE_{\bar{\nu}_{\mu}}}\right|_{ch} dE_{\bar \nu_{\mu}} \cdot T_{\text{eff}}^j, \label{Acc1}
\end{split}
\end{equation}

\par\noindent where $ A^j_{\text{eff}}(E_{\nu_{\mu}})$ and $A^j_{\text{eff}}(E_{\bar \nu_{ \mu}})$ are the effective areas for the $j$-th detector configuration period (see below) as a function of the muon neutrino energy, $ E_{\nu_{\mu}}$, or muon antineutrino energy, $ E_{\bar{\nu}_{\mu}}$, $\left.\frac{d\Phi_{\nu_\mu}}{dE_{\nu_\mu}}\right|_{ch}$ is the signal neutrino spectrum at the position of the detector for the annihilation channel $ch$ (see Equation \ref{channels}), $E_{th}$ is the energy threshold of the detector, $ M_{\text{WIMP}}$ is the WIMP mass and $T_{\text{eff}}^j$ is the effective live time for the $j$-th detector configuration period. The effective area is defined as a 100\% efficient equivalent area which would produce the same event rate as the detector. It is calculated from simulation. Throughout the lifetime of ANTARES the number of available detector lines has changed. The acceptance for the whole lifetime $\bar{\mathcal{A}}$ is calculated as the sum over the acceptances for all detector configuration periods.

The 90\% C.L. sensitivities on the neutrino fluxes are then calculated as

\begin{equation}
 \bar \Phi_{\nu_\mu+\bar \nu_\mu,90\%} =  
\frac{\bar \mu_{90\%}(M_{\text{WIMP}})}{\bar{\mathcal{A}}(M_{\text{WIMP}})},
\end{equation}

\noindent where $\bar \mu_{90\%}$ is the 90\% C.L. sensitivity obtained from the likelihood function. 

\section{Results and discussion}
\label{result}

\begin{figure}[h!]
\centering
\includegraphics[width=0.5\textwidth]{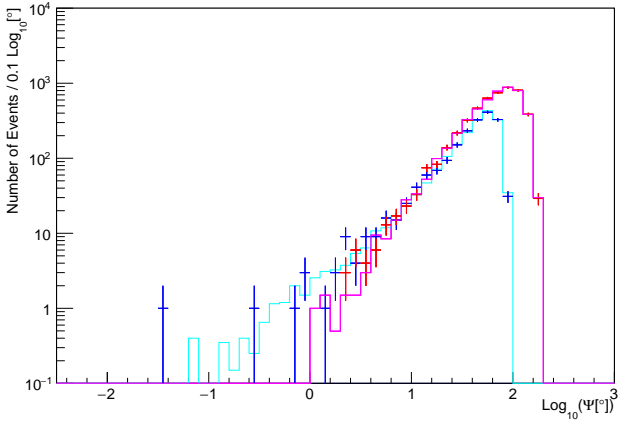}
 \caption{Distribution of the angular distance between reconstructed the track direction of events and the Sun position for the $\Lambda $ Fit  (red and pink) and QFit (blue and cyan) data samples (crosses) compared to the background estimates (histograms). For QFit the x-axis represents the logarithmic difference in zenith angle between event and Sun.}
\label{AAestim}
\end{figure}

In Figure \ref{AAestim} it can be seen that there is no excess of events large enough to be identified as signal by the likelihood function. The median of the PSF used in the likelihood function is for most masses below 2 degrees. The observed TS is used to extract 90\% C.L. upper limits from the absence of signal. However, since the observed value of the TS turns out to be smaller than the median of the TS distribution of pure background for all masses and channels, the sensitivity has been considered as the limit.

\begin{figure}[h!]
\centering
\includegraphics[width=0.5\textwidth]{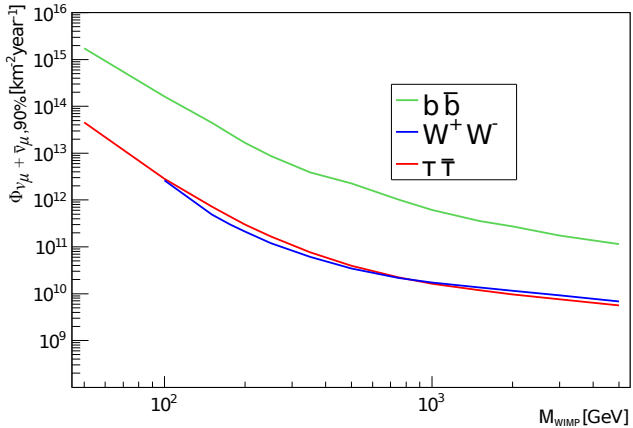}
 \caption{Limits on a neutrino flux coming from the Sun as a function
   of the WIMP masses for the different channels considered.}
\label{FLUX}
\end{figure}

In Figure \ref{FLUX} the limits on the neutrino flux from the Sun as a function of the WIMP mass are shown. In Figure \ref{FLUX} the QFit and $\Lambda $ Fit results are combined.  $\Lambda $ Fit gives the best flux limits in the $W^+ W^-$ decay channel at all WIMP masses, for $M_{WIMP}>100$ GeV in the $\tau^+ \tau^-$ channel and for $M_{WIMP}>250$~GeV in the $b \bar b$ decay channel.

The limit on the total number of neutrinos from WIMP annihilations in the sun per unit of time $C_n$ is calculated by

\begin{equation}
 C_n =  4 \pi d_{\text{Sun},rms}^2 \Phi_{\nu_\mu+\bar \nu_\mu,90\%},
 \end{equation}

\noindent where $\Phi_{\nu_\mu+\bar \nu_\mu,90\%} $ is the limit on the neutrino flux and $d^2_{\text{Sun},rms}$ is the mean squared distance from the detector to the Sun. From this, the annihilation rate is calculated by dividing $C_n$ by the average number of neutrinos per annihilation, as obtained by WIMPSim. The sensitivities on the spin--dependent and spin--independent scattering cross--sections are calculated from this annihilation rate assuming an equilibrium between annihilation and capture via scattering \cite{indirectdm3}. This means that the capture rate is twice as high as the annihilation rate. For the calculation of the capture rate a Maxwellian velocity distribution of the WIMPs with a root mean square velocity of 270 $\text{m}\cdot \text{s}^{-1}$ and a local dark matter density of 0.4 $\text{GeV}\cdot \text{cm}^{-3}$ is assumed \cite{conversion}. Therefore, once the average number of neutrinos per annihilation is known, the annihilation rate and consequently the capture rate and the scattering cross--sections can be calculated. 

\begin{figure}[h!]
\centering
\includegraphics[width=0.48\textwidth]{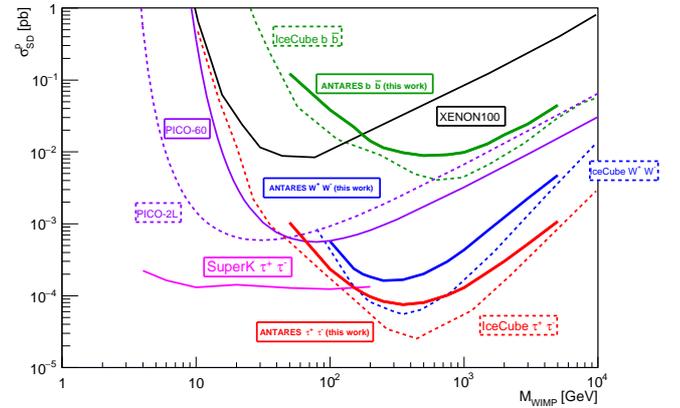}
\caption{Limits on the spin--dependent WIMP--nucleon scattering cross--section as a function of WIMP mass for the $b \bar{b}$, $\tau^+ \tau^-$ and $W^+ W^-$ channels. Limits given by other experiments are also shown: IceCube~\cite{IC_new}, PICO-60~\cite{PICO_60}, PICO-2L~\cite{PICO_L2}, SuperK~\cite{SuperK_low}, XENON100~\cite{XENON_SD}.}
\label{SDCS}
\end{figure}

\begin{figure}[h!]
\centering
\includegraphics[width=0.48\textwidth]{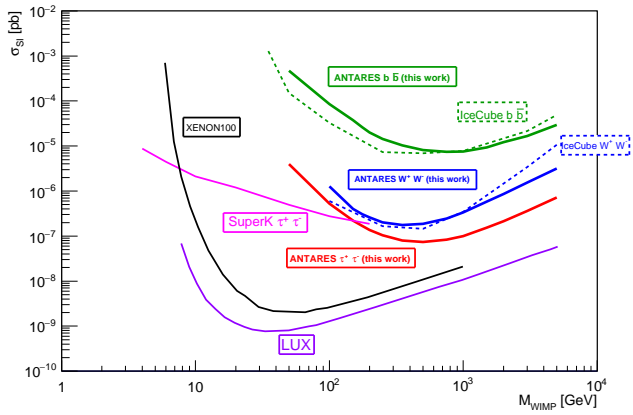}
\caption{Limits on the spin--independent WIMP--nucleon scattering cross--section
as a function of WIMP mass for the different channels considered. Limits given by other experiments are also shown: IceCube~\cite{IC_Sun}, SuperK~\cite{SuperK_low}, LUX~\cite{LUX}, XENON100~\cite{XENON_SI}.}
 \label{SICS}
\end{figure}

All results are shown in comparison to the results of other experiments in Figures \ref{SDCS} and \ref{SICS} and summarised for reference in Table \ref{thetable}. Recently an update on the spin--dependent cross--section limits from the IceCube collaboration has been released ~\cite{IC_new}. These new limits show an improvement of up to a factor of 4 with respect to the previous limits by using the energy information of the events in the likelihood function. In the analysis presented here the inclusion of further event parameters (e.g. $N_{hit}$, $\beta$ and $Q$ in Equations \ref{lik1} and \ref{lik2}) leads to an improvement of a factor of up to 1.7. At WIMP masses of up to a few 100 GeV, the consistent strengthening of the flux limit with increasing WIMP mass (see Figure \ref{FLUX}) determines the behaviour of the cross--section limits. Above a WIMP mass of a few 100 GeV the factor of $M_{WIMP}^{-2}$ in the conversion from neutrino flux to the scattering cross--sections dominates the behaviour of the cross--section limits and causes a rise with the WIMP mass. As a result, the cross-section limits show a minimum at a few 100 GeV. 
 
The possible uncertainties on the background have been circumvented by using time--scrambled data for generating the background function $B$ in the likelihood function. The largest systematic error is an uncertainty of 20\% on the angular acceptance of the PMTs \cite{tical} and leads to a degradation of the detector efficiency (i.e. the acceptance) of 6\% \cite{Ant_dmsun}. This effect has been taken into account for the limits presented here.

\section{Conclusion}

A new analysis searching for a signal of dark matter annihilations in the Sun has been conducted using the ANTARES data from 2007 to 2012. The unblinded data showed no significant excess above the background estimate and 90\% confidence level exclusion limits have been calculated for the three annihilation channels $\text{WIMP} + \text{WIMP} \to b\bar b , W^+ W^- , \tau^+ \tau^-$ and WIMP masses ranging from 50 GeV to 5 TeV. 

\begin{table}
\begin{tabular}[b]{|c|c||c|c|c|}
\hline
$\rm M_{WIMP}$ & & $\Phi_{\nu}$ & $\rm \sigma^{p}_{SD}$ & $\rm \sigma_{SI}$\\
$\rm[GeV]$ & & $\rm [km^{-2}yr^{-1}]$ & $\rm [pb]$ & $\rm [pb]$ \\
\hline 
\hline
50 	& $\rm b \bar{b}$	& $1.86\cdot 10^{15}$ 	& $0.129$				& $4.98\cdot 10^{-4}$ \\
\hline
	& $\tau \bar{\tau}$	& $4.80\cdot 10^{13}$	& $1.10\cdot 10^{-3}$	& $4.23\cdot 10^{-6}$ \\
\hline
100	& $\rm b \bar{b}$	& $1.73\cdot 10^{14}$	& $4.04\cdot 10^{-2}$	& $9.05\cdot 10^{-5}$ \\
\hline
	& $\rm W^+ W^-$	& $2.77\cdot 10^{12}$	& $6.01\cdot 10^{-4}$	& $1.35\cdot 10^{-6}$ \\
\hline
	& $\tau \bar{\tau}$	& $3.02\cdot 10^{12}$	& $2.48\cdot 10^{-4}$	& $5.55\cdot 10^{-7}$ \\
\hline	
150	& $\rm b \bar{b}$	& $4.78\cdot 10^{13}$	& $2.36\cdot 10^{-2}$	& $4.00\cdot 10^{-5}$ \\
\hline
	& $\rm W^+ W^-$	& $5.23\cdot 10^{11}$	& $2.52\cdot 10^{-4}$	& $4.26\cdot 10^{-7}$ \\
\hline	
	& $\tau \bar{\tau}$	& $7.69\cdot 10^{11}$	& $1.39\cdot 10^{-4}$	& $2.35\cdot 10^{-7}$ \\
\hline	
176	& $\rm b \bar{b}$	& $2.70\cdot 10^{13}$	& $1.81\cdot 10^{-2}$	& $2.77\cdot 10^{-5}$ \\
\hline
	& $\rm W^+ W^-$	& $3.18\cdot 10^{11}$	& $2.12\cdot 10^{-4}$	& $3.24\cdot 10^{-7}$ \\
\hline	
	& $\tau \bar{\tau}$	& $4.67\cdot 10^{11}$	& $1.15\cdot 10^{-4}$	& $1.77\cdot 10^{-7}$ \\
\hline	
200	& $\rm b \bar{b}$	& $1.76\cdot 10^{13}$	& $1.51\cdot 10^{-2}$	& $2.13\cdot 10^{-5}$ \\
\hline
	& $\rm W^+ W^-$	& $2.25\cdot 10^{11}$	& $1.95\cdot 10^{-4}$	& $2.71\cdot 10^{-7}$ \\
\hline	
	& $\tau \bar{\tau}$	& $3.19\cdot 10^{11}$	& $1.10\cdot 10^{-4}$	& $1.43\cdot 10^{-7}$ \\
\hline
250	& $\rm b \bar{b}$	& $8.75\cdot 10^{12}$	& $1.15\cdot 10^{-2}$	& $1.43\cdot 10^{-5}$ \\
\hline
	& $\rm W^+ W^-$	& $1.25\cdot 10^{11}$	& $1.72\cdot 10^{-4}$	& $2.15\cdot 10^{-7}$ \\
\hline
	& $\tau \bar{\tau}$	& $1.75\cdot 10^{11}$	& $8.82\cdot 10^{-5}$	& $1.10\cdot 10^{-7}$ \\
\hline	
350	& $\rm b \bar{b}$	& $4.11\cdot 10^{12}$	& $1.03\cdot 10^{-2}$	& $1.09\cdot 10^{-5}$ \\
\hline
	& $\rm W^+ W^-$	& $6.46\cdot 10^{10}$	& $1.77\cdot 10^{-4}$	& $1.88\cdot 10^{-7}$ \\
\hline	
	& $\tau \bar{\tau}$	& $8.03\cdot 10^{10}$	& $7.95\cdot 10^{-5}$	& $8.44\cdot 10^{-8}$ \\
\hline	
500	& $\rm b \bar{b}$	& $2.37\cdot 10^{12}$	& $9.36\cdot 10^{-3}$	& $8.64\cdot 10^{-6}$ \\
\hline
	& $\rm W^+ W^-$	& $3.67\cdot 10^{10}$	& $2.13\cdot 10^{-4}$	& $1.98\cdot 10^{-7}$ \\
\hline	
	& $\tau \bar{\tau}$	& $4.20\cdot 10^{10}$	& $8.48\cdot 10^{-5}$	& $7.82\cdot 10^{-8}$ \\
\hline	
750	& $\rm b \bar{b}$	& $1.08\cdot 10^{12}$	& $9.68\cdot 10^{-3}$	& $7.95\cdot 10^{-6}$ \\
\hline
	& $\rm W^+ W^-$	& $2.29\cdot 10^{10}$	& $3.16\cdot 10^{-4}$	& $2.59\cdot 10^{-7}$ \\
\hline	
	& $\tau \bar{\tau}$	& $2.36\cdot 10^{10}$	& $1.07\cdot 10^{-4}$	& $8.82\cdot 10^{-8}$ \\
\hline	
1000	& $\rm b \bar{b}$	& $6.52\cdot 10^{11}$	& $1.04\cdot 10^{-2}$	& $8.03\cdot 10^{-6}$ \\
\hline
	& $\rm W^+ W^-$	& $1.83\cdot 10^{10}$	& $4.59\cdot 10^{-4}$	& $3.55\cdot 10^{-7}$ \\
\hline	
	& $\tau \bar{\tau}$	& $1.72\cdot 10^{10}$	& $1.37\cdot 10^{-4}$	& $1.06\cdot 10^{-7}$ \\
\hline
1500	& $\rm b \bar{b}$	& $3.79\cdot 10^{11}$	& $1.37\cdot 10^{-2}$	& $9.95\cdot 10^{-6}$ \\
\hline
	& $\rm W^+ W^-$	& $1.44\cdot 10^{10}$	& $8.47\cdot 10^{-4}$	& $6.15\cdot 10^{-7}$ \\
\hline
	& $\tau \bar{\tau}$	& $1.26\cdot 10^{10}$	& $2.24\cdot 10^{-4}$	& $1.63\cdot 10^{-7}$ \\
\hline
2000	& $\rm b \bar{b}$	& $2.88\cdot 10^{11}$	& $1.82\cdot 10^{-2}$	& $1.28\cdot 10^{-5}$ \\
\hline
	& $\rm W^+ W^-$	& $1.21\cdot 10^{10}$	& $1.30\cdot 10^{-3}$	& $9.17\cdot 10^{-7}$ \\
\hline	
	& $\tau \bar{\tau}$	& $1.03\cdot 10^{10}$	& $3.20\cdot 10^{-4}$	& $2.25\cdot 10^{-7}$ \\
\hline	
3000	& $\rm b \bar{b}$	& $1.82\cdot 10^{11}$	& $2.60\cdot 10^{-2}$	& $1.78\cdot 10^{-5}$ \\
\hline
	& $\rm W^+ W^-$	& $9.73\cdot 10^{9}$		& $2.44\cdot 10^{-3}$	& $1.63\cdot 10^{-6}$ \\
\hline	
	& $\tau \bar{\tau}$	& $8.01\cdot 10^{9}$		& $5.57\cdot 10^{-4}$	& $3.81\cdot 10^{-7}$ \\
\hline	
5000	& $\rm b \bar{b}$ 	& $1.20\cdot 10^{11}$	& $4.71\cdot 10^{-2}$	& $3.15\cdot 10^{-5}$ \\
\hline
	& $\rm W^+ W^-$	& $7.25\cdot 10^{9}$		& $5.02\cdot 10^{-3}$	& $3.36\cdot 10^{-6}$ \\
\hline
	& $\tau \bar{\tau}$	& $5.02\cdot 10^{9}$		& $1.13\cdot 10^{-3}$	& $7.62\cdot 10^{-7}$ \\
\hline	
\end{tabular}
\caption{Upper limits to neutrino flux, spin--dependent and spin--independent cross-section for different annihilation channels and WIMP masses. Limits for the $W^+ W^-$ channel cannot be produced for WIMP masses below the mass of the W boson.
\label{thetable}}
\end{table}

\section*{Acknowledgements}

The authors acknowledge the financial support of the funding agencies:
Centre National de la Recherche Scientifique (CNRS), Commissariat \`a
l'\'ener\-gie atomique et aux \'energies alternatives (CEA),
Commission Europ\'eenne (FEDER fund and Marie Curie Program),
Institut Universitaire de France (IUF),
IdEx program and UnivEarthS Labex program at Sorbonne Paris Cit\'e
(ANR-10-LABX-0023 and ANR-11-IDEX-0005-02), R\'egion
\^Ile-de-France (DIM-ACAV), R\'egion Alsace (contrat CPER), R\'egion
Provence-Alpes-C\^ote d'Azur, D\'e\-par\-tement du Var and Ville de La
Seyne-sur-Mer, France; Bundesministerium f\"ur Bildung und Forschung
(BMBF), Germany; Istituto Nazionale di Fisica Nucleare (INFN), Italy;
Stichting voor Fundamenteel Onderzoek der Materie (FOM), Nederlandse
organisatie voor Wetenschappelijk Onderzoek (NWO), the Netherlands;
Council of the President of the Russian Federation for young
scientists and leading scientific schools supporting grants, Russia;
National Authority for Scientific Research (ANCS), Romania; 
Mi\-nis\-te\-rio de Econom\'{\i}a y Competitividad (MINECO),
Severo Ochoa Centre of Excellence, MultiDark Consolider, Prometeo and
Grisol\'{\i}a programs of Generalitat Valenciana, Spain; 
Agence de  l'Oriental and CNRST, Morocco. We also acknowledge 
the technical support of Ifremer, AIM and Foselev Marine for the sea 
operation and the CC-IN2P3 for the computing facilities.

\appendix

\section*{References}





\bibliography{manuscript}

\end{document}